\documentclass[preprint,aps,nofootinbib]{revtex4}
\pdfoutput=1
\usepackage{graphicx}% Include figure files
\usepackage{epstopdf}    % nado dobavit, inache pdf ne budet
\usepackage{dcolumn}% Align table columns on decimal point
\usepackage{amsfonts}
\usepackage{amsmath}
\usepackage{amssymb}
\usepackage{xcolor}
\usepackage{bm}% bold math
\usepackage{color}
\usepackage{comment}
\usepackage{epsf}
\usepackage{float}
\usepackage{enumitem}

\usepackage[colorlinks=true,citecolor=blue]{hyperref}
\hypersetup{colorlinks=true,citecolor=blue,linkcolor=red,urlcolor=blue}

\counterwithin{equation}{section}

\newcolumntype{L}{>{\centering\arraybackslash}m{3cm}}
	
	\newcommand{\bqa}{\begin{eqnarray}}
	\newcommand{\eqa}{\end{eqnarray}}
	\newcommand{\bwt}{\begin{widetext}}
	\newcommand{\ewt}{\end{widetext}}

\newcommand{\edc}{\end{document}}
\newcommand{\bb} {\begin{thebibliography}}
\newcommand{\eb}{\end{thebibliography}}
\newcommand{\bi}[1]{\bibitem{#1}}
\newcommand{\bc}{\begin{center}}
\newcommand{\ec}{\end{center}}

\newcommand{\be}{\begin{equation}}
\newcommand{\ee}{\end{equation}\normalsize}
\newcommand{\bea}{\begin{eqnarray}}
\newcommand{\eea}{\end{eqnarray}}
\newcommand{\ba}{\begin{array}{l}   }
\newcommand{\lab}[1]{\label{#1}}
\newcommand{\ea}{\end{array}}

\newcommand{\dsfrac}{\displaystyle\frac}
\newcommand{\ds} {\displaystyle}

\newcommand{\re}[1]{(\ref{#1})}

\newcommand{\ci}{\cite}

\newcommand{{\vergul}}{  ,}

\newcommand{\veps}{\varepsilon }

\newcommand{\rqavs}{\right )}
\newcommand{\lqavs}{\left (}

\renewcommand{\theequation}{\arabic{section}.\arabic{equation}}

\begin{document}

	%\doublespace
\title{Critical behavior of Tan's contact for bosonic systems with a fixed chemical potential}

	\author{Abdulla Rakhimov$^1$, Tolib Abdurakhmonov$^1$,  B. Tanatar$^2$ }
	%\email{rakhimovabd@yandex.ru,  t.aburakhmonov@mail.ru, tanatar@fen.bilkent.edu.tr}
	\affiliation{$^1$Institute of Nuclear Physics, Tashkent 100214, Uzbekistan \\
	$^2$Department of Physics, Bilkent University, Bilkent, 06800 Ankara, Turkey}

		\date{\today}
		
		\begin{abstract}
The temperature dependence of Tan's contact parameter $C$ and its derivatives for spin gapped quantum magnets are investigated. We use the paradigm of Bose-Einstein condensation (BEC) to describe the low temperature properties of 
quasiparticles in the system known as triplons. Since the number of particles and the condensate fraction are 
not fixed we use the $\mu VT$ ensemble to calculate the thermodynamic quantities. The interactions are treated 
at the Hartree-Fock-Bogoliubov approximation level. We obtained the temperature dependence of $C$ and its
derivative with respect to temperature and applied magnetic field both above and below  $T_c$ of the 
phase transition from the normal phase to BEC. 
We have shown that $C$ is regular, while its derivatives are discontinuous at $T_c$ in accordance
with Ehrenfest's classification of phase transitions. Moreover, we have found 
a sign change in $\partial C/\partial T$ close to the critical temperature. As to the quantum critical point, $C$ and its derivatives are regular as a function of the control parameter $r$, which induces the quantum phase transition.
At very low temperatures,  one may evaluate  $C$ simply from the expression 
$C=m^2\mu^2/{\bar a}^{4}$, where the only parameter effective mass of quasiparticles  should be estimated.
We propose a method for measuring of Tan's contact for spin gapped dimerized magnets.

  		\end{abstract}
\maketitle

%\newpage
\section{Introduction}
%\label{sec:1}
Some years ago, Tan derived a set of exact relations that link the short distance large-momentum correlations to the bulk thermodynamic properties of a fermionic system with short-range inter-particle interactions \ci{Tan1, Tan2, Tan3}. They are connected by a single coefficient $C$, referred to as the integrated contact intensity or ``contact". Further, Tan's ideas were developed \ci{Br8, Br9} and extended to Bose systems also
\ci{Br10,Lang11,Werner1,Combescot13}. It has been established that
Tan's contact measures the density of pairs at short distances and determines the exact large-momentum
or high frequency behavior of various physical observables. It serves as an important 
quantity to characterize strongly interacting many-body systems \ci{Pitbook14}. 

Experimentally, Tan's relations have been confirmed both for fermions (especially for quasi-1D systems) \ci{kuhle2011, sagi2012, stewart2010} and Bose gases \ci{chang2016, Fletcher2016, makotyu2014, wild2012, zou2021}.

Although, there are several theoretical studies of contact of low dimensional Fermi systems even at 
finite temperatures \ci{Hoffman,yuPRA80,Matveeva,Br8,brateen2013,vigndofermi,Patu}, the temperature dependence of bosonic contact in three dimensions is mostly unknown. 

Studying the temperature dependence of contact parameter gives an opportunity to know its critical behavior near phase transitions. For example, Chen {\it et al}. \ci{chencrit} have shown {that contact and its derivatives are uniquely determined by the universality class of the phase transition for fermions in the critical region.}
In a system of bosons, critical behavior of thermodynamic quantities have some specifics due to Bose-Einstein condensation (BEC) at low temperatures. For instance, the specific heat is discontinuous at the critical temperature $T_c$, and the Gr{\"u}neisen parameter changes its sign \ci{garst,ourMCE}.

The main goal of the present work is to study the temperature dependence of Tan's contact for bosons, as well as its derivatives, to draw conclusions about their critical behavior near $T_c$  and close to the quantum critical
point (QCP) of the quantum phase transition (QPT).

It is well known that physical systems at equilibrium are studied in statistical mechanics through statistical 
ensembles. For example, a system {that} the canonical ensemble exchanges only energy with the surroundings 
is called $NVT$, where $N$ is the number of atoms of a gas in the volume $V$ at temperature $T$. 
On the other hand,
a grand canonical ensemble {also makes it possible for particles to exchange}, 
with the parameter $\mu$, which may be 
interpreted as the chemical potential, and is called $\mu VT$. In the former case, $N$ or $\rho=N/V$ is fixed 
and the chemical potential is determined by the equation of state, while in the latter case $\mu$ is fixed as an 
input parameter which defines number of particles through a thermodynamic equation \ci{Marzolino}. 
Tan's relations including the contact have been thoroughly studied for $NVT$ ensembles, {however, the investigations of contact for a $\mu VT$ ensemble is still missing. In this work, we choose 
a system} of bosonic quasiparticles introduced in bond operator formalism \ci{Sachdev,OurAnn2} to describe 
the singlet-triplet excitations in spin gapped magnetic materials \ci{Zapf}. It is well established that low temperature
properties of such a class of quantum magnets could be described within the paradigm of Bose-Einstein
condensation of these quasiparticles {called} triplons. Therefore,  one may conclude that,  at low 
temperature, thermodynamic properties of such materials are {mainly determined} (but not only) by the 
condensation (and depletion) of triplons \ci{Zapf,ourMCE,ourJT,ourCharac}. Triplon concept has found 
application in other spin-gapped models as well \cite{kumar}.

There is one more reason of our choice of spin gapped quantum magnets to study Tan's contact. 
In such systems, the number of particles $N$ and the condensate fraction $N_0/N$ can be directly evaluated by measuring the uniform magnetization $M$ and staggered magnetization $M_{\perp}$ respectively, 
as $N=M/g\mu_B$, $N_0^2=2M_{\perp}^2/(g\mu_B)^2$
\ci{delamorre}, where $g$ is the electron Land{\'e} factor and $\mu_B$ is the  Bohr magneton. 
Evidently, the final analytic expression for $C$ will include $M$ and $M_{\perp}$, making the contact parameter
to be easily measured.

In practice, Tan's contact for bosons with zero range interaction may be evaluated by any of following Tan's relations \ci{Pitbook14}. {For example,}
\begin{enumerate}
\item By the asymptotic behavior of momentum distribution $n_k$, 
\be
C=\lim_{k\rightarrow \infty}k^4 n_k\, .
\ee
\item By Tan's sweep theorem as 
\be
C=\frac{8\pi ma^2}{V} \ 
\begin{cases}
%\left(\dsfrac{\partial E}{\partial a}\right) \ \ \ & \text{$T=0$} \\
\left(\dsfrac{\partial F}{\partial a}\right)_{N,T} \ \ \ & \text{for $NVT$ ensembles}\, , \\
\left(\dsfrac{\partial\Omega}{\partial a}\right)_{\mu,T} \ \ \ & \text{for ${\mu}VT $ensembles}\, ,
\end{cases}
\label{Cdef2}
\ee
where  $F$ and $\Omega$ are the free energy and the grand canonical potential, respectively \ci{bouchul21}; $a$ is the scattering length and $m$ is the effective mass of the particle.
\item By using Hellman-Feynman theorem \ci{Br8}
\be
C=\frac{16\pi^2 a^2}{V}\int d\vec{r}\langle\psi^{\dagger}(\vec{r})\psi^{\dagger}(\vec{r})\psi(\vec{r})\psi(\vec{r})\rangle\, .
\ee
\end{enumerate}
In our previous paper \ci{ourTan1}, it has been shown that if one uses mean field theory (MFT), evaluation of $C$ 
from Eq.\,(\ref{Cdef2}) will be the most convenient and reliable. {Consequently, we study $C$}
of a triplon gas at finite temperature in the framework of MFT, namely in the Hartree-Fock-Bogoliubov (HFB)
approximation \ci{ourTan13}.
{We stress here that below we shall discuss $C$ only for homogeneous systems. For inhomogeneous systems,
especially at very low temperature, one may also use the Gross-Pitaevskii equation to study the
dynamics of the condensate \ci{wangpra81,wangpra84} as well as  that of the Tan's contact.}

{For convenience, we adopt units such that $\hbar=1$, $k_B=1$ and $V=1$ in the following text}. In these units 
$C$ is dimensionless.
In natural units $C$ may be obtained further by dividing it by ${\bar a}^4$, getting it in 
(length)$^{-4}$ where ${\bar a}$
is the average lattice parameter, ${\bar a}=V^{1/3}$, and $V$ is the unit cell volume of the crystal.
 
{This article is organized as follows}. In Section II, we outline main equations for a triplon gas at finite temperature in the HFB approximation, which will be used to evaluate the Tan's contact and its derivatives
in Section III. Our discussions and conclusions will be presented in Section IV.

\section{Triplon density in the HFB approximation}

We start with the effective Hamiltonian of triplons as a non-ideal Bose gas with contact repulsive interaction
\be
{\ds\hat H}=\int d\vec{r}\left[\psi^{\dagger}(\vec{r})(\hat{K}-\mu)\psi(\vec{r})+\frac{U}{2}(\psi^{\dagger}(\vec{r}) \psi(\vec{r}))^2 \right]
\label{Ham}
\ee
where $\psi(\vec{r})$ is the bosonic field operator, $U=4\pi a/m$ is the interaction strength, 
and $\hat{K}$ is the kinetic energy operator which defines the bare triplon dispersion $\varepsilon_{\vec{k}}$ 
in momentum space. Since the triplon BEC occurs in solids, the integration is performed over the unit cell of the
crystal with the corresponding momenta defined in the first Brillouin zone \ci{ouraniz1}. The parameter $\mu$ characterizes an additional direct contribution to the triplon energy due to the external field
\be
\mu=g \mu_B H-\Delta       
\label{chempot}
\ee
and it can be interpreted as the chemical potential of the $S_z=-1$ triplons
\ci{Giamarchi}.
%=,ouraniz8]. 
In Eq.\,(\ref{chempot}),  $\Delta=g\mu_B H_c$, 
is the spin gap separating the ground-state from the lowest energy triplet excitation
and $H_c$ is the critical external magnetic field \cite{grundmann}. {The effective 
Hamiltonian} ${\hat H}$ has the following
free parameters characterizing a given material: $g$, $H_c$, $U$ and possibly, an effective triplon mass $m$
for the case of a simple bare dispersion given as $\veps_{\vec k}={\vec k}^2/2m$.\footnote{
In general, a realistic  $\veps_{\vec k}$ includes several other parameters \ci{misguich}.}

In general the Hamiltonian in Eq.\,(\ref{Ham}) is invariant under global $U(1)$ gauge transformation $\psi(\vec{r})\rightarrow e^{i\alpha}\psi(\vec{r})$ with {a real number $\alpha$}. However, this symmetry is broken in the condensate phase, where $T<T_c$, and it is restored in the normal phase, $T\geqslant T_c$.

This transition temperature $T_c$, which corresponds to the vanishing of the condensate density 
$\rho_0(T_c)=0$ may be calculated from the following equation
\be
\sum_k \frac{1}{e^{\varepsilon_k/T_c}-1}=\frac{\mu}{2U}\equiv \rho_c\, .
\label{Tc}
\ee
Note that for existing quantum magnets with a spin gap $T_c$ is rather large, $T_c\approx 2$\,K, in contrast to the critical temperature of BEC in atomic gases, where the number of particles is fixed
and the critical temperature is of the order of nanokelvins. However, in the $\mu VT$ ensemble 
under discussion, the particle density depends on the external magnetic field through the chemical potential $\mu$ 
as 
\be
\ba
\rho(T \ge T_c)=\ds\sum_k \frac{1}{e^{\beta\omega_k}-1}, \quad \quad 
\omega_k=\varepsilon_k-\mu+2U\rho\, ,
\label{rhobig}
\ea
\ee
in the normal phase. In the BEC phase, an explicit expression for $\rho(T<T_c)$ strongly depends on the chosen
version of an approximation. For example, in the HFB approximation, which is employed here, the particle 
density and condensate fraction are given by the following set of equations \ci{ourAnn}
\begin{eqnarray}
\begin{aligned}
 \rho(T < T_c)=\rho_0+\rho_1=\frac{\mu+\mu_{eff}}{2U}\, , \\
 \mu_{eff}=\mu+2U(\sigma-\rho_1)\, , \\
 \rho_1=\frac{1}{2}\sum_k\left[\frac{\coth(\beta E_k/2)(\varepsilon_k+\mu_{eff})}
{E_k}-1\right]\, , \\
 \sigma=-\frac{\mu_{eff}}{2}\sum_k\frac{\coth(\beta E_k/2)}{E_k}\, ,
\label{MainEqs}
\end{aligned}
\end{eqnarray}
where $\rho_1$ and $\sigma$ correspond to the density of non-condensed particles and anomalous 
density, respectively. 
The hyperbolic function in Eq.\,\re{MainEqs} can be represented also as  $\coth(\beta x/2)=1/2+f_B(x)$, 
where $f_B(x)=1/(\exp(\beta x)-1)$ is the Bose distribution function.

Note that neglecting $\sigma$ leads to an unexpected cusp in magnetization \ci{ourPRB81}. 
In Eq.\,(\ref{MainEqs}), $E_k$ is the dispersion of quasiparticles (bogolons) given as $E_k=\sqrt{\varepsilon_k}
\sqrt{\varepsilon_k+2\mu_{eff}}$ with the speed of sound $c=\sqrt{\mu_{eff}/m}$, where $m$ has the meaning of the triplon effective mass, corresponding to  the limit of small momenta $\varepsilon_k \approx \vec{k}^2/2m$. The typical value of $m$, used in the literature, 
\ci{Zapf} is  rather small, $m\approx 0.02$\,K.

\section{Tan's contact for a triplon gas}

As it is pointed out in the Introduction, for the $\mu VT$ ensemble it is convenient to calculate $C$ using
\be
C=2U^{2}m^{2}\left(\frac{\partial\Omega}{\partial U}\right)_{T,\mu}\, ,
\label{CDeRho}
\ee
where the grand thermodynamic potential $\Omega$ has the following total derivative \ci{Pitbook14,ourJT,ourCharac}
\begin{equation}
d\Omega=-SdT-PdV-Nd\mu-MdH+\frac{C da}{8\pi m a^2}\, ,
\label{domega}
\end{equation}
in which $S$ is the entropy and $P$ is the pressure.
Note that the relation between the magnetization $M$ and the triplon density may be directly
obtained from Eq.\,(\ref{domega}) as $M=g\mu_B\rho$.
Moreover, using Eq.\,(\ref{domega}) one may find useful expressions for the derivatives of $C$ as
\bea
\begin{aligned}
  \left(\frac{\partial C}{\partial T}\right)=-2U^2m^2\left(\frac{\partial S}{\partial U}\right)\, , \\
  \lqavs\frac{\partial C}{\partial \mu}\rqavs=
\dsfrac{1}{g\mu_B}\left(\frac{\partial C}{\partial H}\right) =
-2U^2m^2\left(\frac{\partial \rho}{\partial U}\right)\, .
\label{cdef}
\end{aligned}
\eea

%%%%%%%%%%%%%%%%%%%%  added NEW TEXT
%BEGIN
It is well known that {from BEC to a normal phase transition} is a second order phase transition,
in which the entropy $S(T,H,a)$ is continuous across the critical temperature i.e., 
\be
dS(T=T_c^-)=dS(T=T_c^+)=\left ( \dsfrac{\partial S}{\partial T}\right )_{H,a}dT+
\left ( \dsfrac{\partial S}{\partial H}\right )_{T,a}dH+\left ( \dsfrac{\partial S}{\partial a}\right )_{H,T}da\, .
\lab{ds}
\ee
Therefore, using Eqs.\,\re{domega}-\re{ds} leads to an extended Ehrenfest relation:
\be
T_{c}^{-1}\Delta  [C_H]=-\Delta\left[\left ( \dsfrac{\partial M}{\partial T}\right )_{H,U}\right]\lqavs\dsfrac {dH}{dT}
\rqavs+
\dsfrac{1}{2U^2m^2}\Delta\left[\left ( \dsfrac{\partial C}{\partial T}\right )_{H,U}\right]\lqavs\dsfrac{dU}{dT}\rqavs
\lab{eren}
\ee
where $C_H=T(\partial S/\partial T)$ is the heat capacity and $\Delta [f]\equiv f(T=T_c^-)-f(T=T_c^+)$ is the
jump in the function $f$ at $T=T_c$.\footnote{A similar relation for $NVT$ ensembles will be 
$T_{c}^{-1}\Delta [C_P]=-\Delta\left[\left ( \dsfrac{\partial P}{\partial T}\right )_{V,U}\right]
\lqavs\dsfrac{dV}{dT}\rqavs+
\dsfrac{1}{2U^2m^2}\Delta\left[\left ( \dsfrac{\partial C}{\partial T}\right )_{V,U}\right]\lqavs\dsfrac{dU}{dT}\rqavs$
when the entropy is considered as a function of temperature, volume and the strength of the contact interaction.} 

%%%%%%%%%%%%%%%%%%%%%%%%  end  

In the following we discuss the normal and BEC phases, separately.

\subsection{Normal phase ($T\ge T_c$)}

In the normal phase $\Omega$ is given by
\be
\Omega(T \ge T_c)=-U\rho^2+T\sum_k \ln(1-e^{\beta\omega_k}),
\label{omegaBig}
\ee
where $\rho$ and $\omega_k$ are defined in Eq.\,(\ref{rhobig}).
Now, taking the derivative of $\Omega$ and using Eq.\,(\ref{rhobig}) one may easily find
\be
\lqavs\frac{\partial\Omega}{\partial U} \rqavs_{T \ge T_c}=-\rho^2-2U\rho\rho'+(2\rho+2U\rho')\sum_k\frac{1}{e^{\beta\omega_k}-1}=\rho^2\, ,
\label{difomUBg}
\ee
and hence
\be
C(T \ge T_c)=2U^{2}m^{2}\rho^{2}=\frac{2U^{2}m^{2}M^{2}}{(g\mu_{B})^{2}}\, .
\label{Cbig}
\ee
The derivatives of $C$ can be evaluated using Eq.\,(\ref{cdef}) or directly from Eq.\,\re{Cbig} as
\be
\begin{aligned}
\lqavs\frac{\partial C}{\partial T}\rqavs_{T\ge T_c}=4U^2m^2\rho\lqavs \frac{\partial\rho}{\partial T}
\rqavs 
=-2m^2U^2\left(\frac{\partial S}{\partial U} \right)\, , \\
\lqavs\frac{\partial C}{\partial\mu}\rqavs_{T \ge T_c}=4U^2m^2\rho\lqavs \frac{\partial\rho}{\partial\mu}
\rqavs =-2m^2U^2\left(\frac{\partial\rho}{\partial U} \right)\, ,
\label{difCTCmu}
\end{aligned}
\ee
where \ci{ourMCE}
%link MCE

%================================
%maqola 7
\be
\begin{aligned}
\lqavs \frac{\partial S}{\partial U}\rqavs_{T \ge  T_c}=\frac{2\rho\beta I_1}{1-2I_2}, \ \ \
\lqavs \frac{\partial\rho}{\partial U}\rqavs_{T\ge T_c}=\frac{2\rho I_2}{U[1-2I_2]}, \\
 \\
\label{drodSUBg} 
\end{aligned}
\ee
and  the integrals $I_1$ are $I_2$ {are} given in the Appendix.
Note that Eqs.\,(\ref{difCTCmu}) lead to the following relations for the susceptibility, $\chi=(\partial M/\partial H)$ 
in the normal phase
\be
\begin{aligned}
\lqavs \frac{\partial C}{\partial H}\rqavs_{T \ge T_c}=\frac{4U^2m^2}{(g\mu_B)^2} \lqavs M \chi \rqavs_{T \ge T_c}\, , \\
\chi ({T \ge T_c})=-\frac{(g\mu_B)^2}{2M}\left(\frac{\partial M}{\partial U}\right) _{T \ge T_c}.
\label{chi1}
\end{aligned}
\ee

\subsection{Condensed phase ($T<T_c$)}
	In the condensed phase the thermodynamic potential is given by \ci{ourJT,ourCharac,OurAnn2}
\be
\Omega(T<T_c)=\frac{U\rho_0^2}{2}-\mu\rho_0-\frac{U}{2}(2\rho_1^2+\sigma^2)+\frac{1}{2}\sum_k(E_k-\varepsilon_k-\mu_{eff})+T\sum_k \ln(1-e^{-\beta E_k})
\label{omegBEC}
\ee
with $\mu_{eff}$, $\rho_1$ and $\rho$ are given by Eqs.\,(\ref{MainEqs}), $\rho_0=\rho-\rho_1=\mu_{eff}/U-\sigma$ and $E_k=\sqrt{\varepsilon_k}\sqrt{\varepsilon_k+2\mu_{eff}}$. For simplicity, one may start with the well-known Bogoliubov approximation, where the depletion and anomalous density are completely neglected. For atomic gases, this leads to the Gross-Pitaevskii equation which is successfully used in a lot of systems \ci{Pitbook14}. In the present case Bogoliubov approximation corresponds to $\rho_0\approx\rho$, $\mu_{eff}\approx\mu=U\rho$ and hence
\be
\Omega_{\rm Bogoliubov}(T<T_c)=-\frac{\mu^2}{2U}+\frac{1}{2}\sum_k[E_k-\varepsilon_k-\mu]+T\sum_k \ln(1-e^{-E_k\beta})\, .
\label{Ombog}
\ee
%---------------------------
%
%------------------------------
%\newpage
Since in $\mu VT$ ensembles the chemical potential does not depend on the inter-particle interaction, i.e., 
$(\partial\mu/ \partial U)=0$, one immediately obtains from Eq.\,(\ref{Ombog}) Tan's contact in the Bogoliubov
approximation as\footnote{In natural units $C_{0}=m^2\mu^2/\bar{a}^4$, where 
 the average lattice parameter $\bar{a}$ can be taken from Table 1.}
\be
C_{0}=m^2\mu^2\, .
\label{Cbog}
\ee
From this equation, it is seen that in this simple approximation Tan's contact does not depend on temperature, 
which means that in $\mu VT$ ensembles Bogoliubov approximation is not applicable to finite temperatures. 
The quality of this approximation at $T=0$ is tested below.
In Figs.\,1, Tan's contact as a function of the magnetic field is presented for two different compounds 
TlCuCl$_3$ (Fig.\,1a) and Ba$_3$Cr$_2$O$_8$ (Fig.\,1b) in natural units, i.e., in \AA$^{-4}$. It is seen that for both compounds, the difference between the exact HFB (solid lines) and simple Bogoliubov approximation
(dashed lines) is rather small, especially for small chemical potentials. Remarkably, the contact given 
by Eq.\,(\ref{Cbog}) does not include the scattering length of triplon interaction but includes only the effective 
mass as an internal microscopic parameter of the system. This is why the contact for TlCuCl$_3$ at the 
same magnetic field e.g., at $H=13.5$\,T, is much smaller than that for Ba$_3$Cr$_2$O$_8$. In fact, 
as it is seen from Table 1 the ratio of effective masses is of the order of 
$m$({Ba$_3$Cr$_2$O$_8$})/$m$({TlCuCl$_3$})$\approx 10$.

%%%%%%%%%%%%%%%FIGURE 1 %%%%%%%%%%%%%%%%%%%%%%%%%%
%Here Figure 1
\begin{figure}[h!]
\begin{minipage}[H]{0.49\linewidth}
\center{\includegraphics[width=1.25\linewidth]{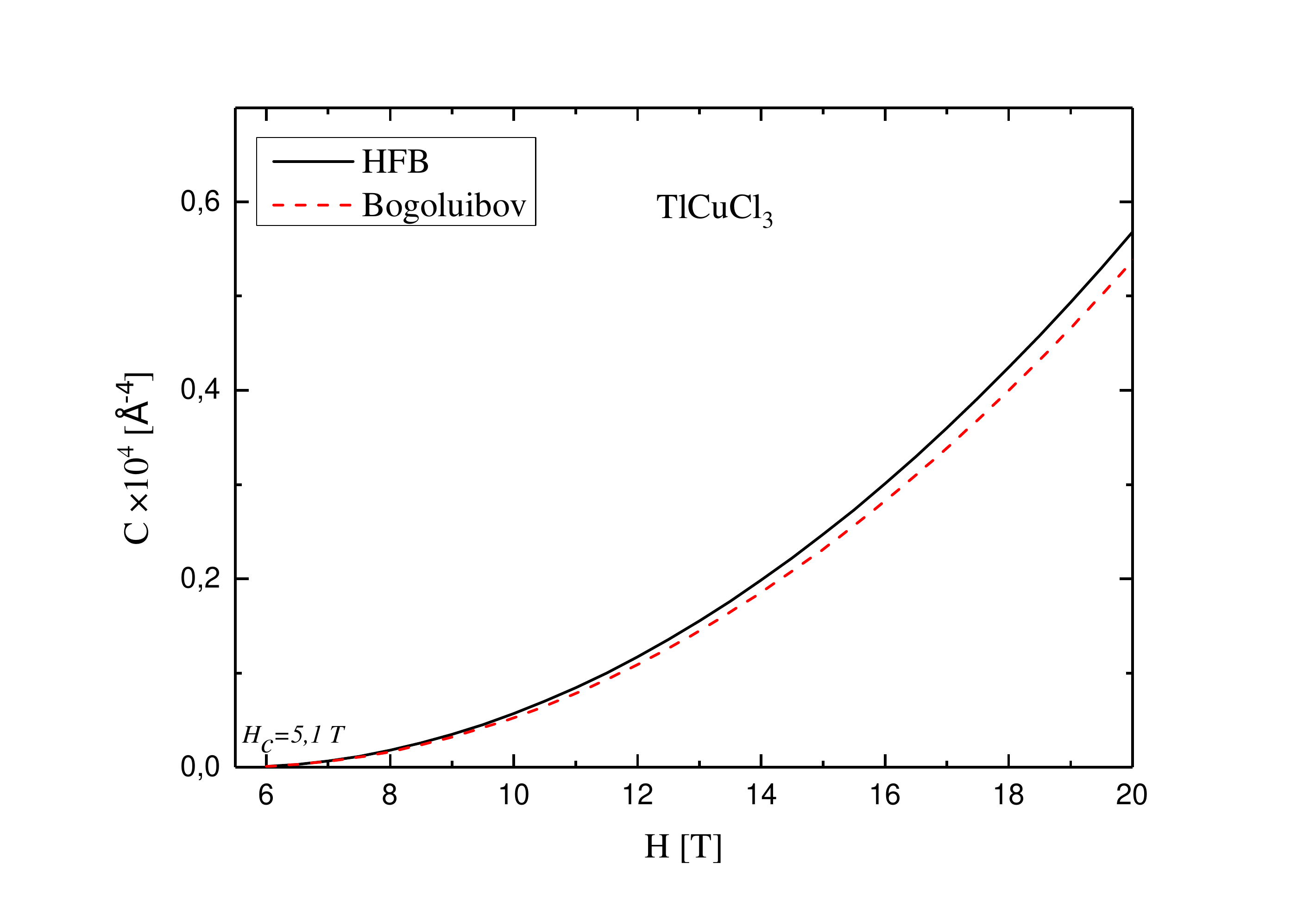} \\ a)}
\end{minipage}
\hfill
\begin{minipage}[H]{0.49\linewidth}
\center{\includegraphics[width=1.25\linewidth]{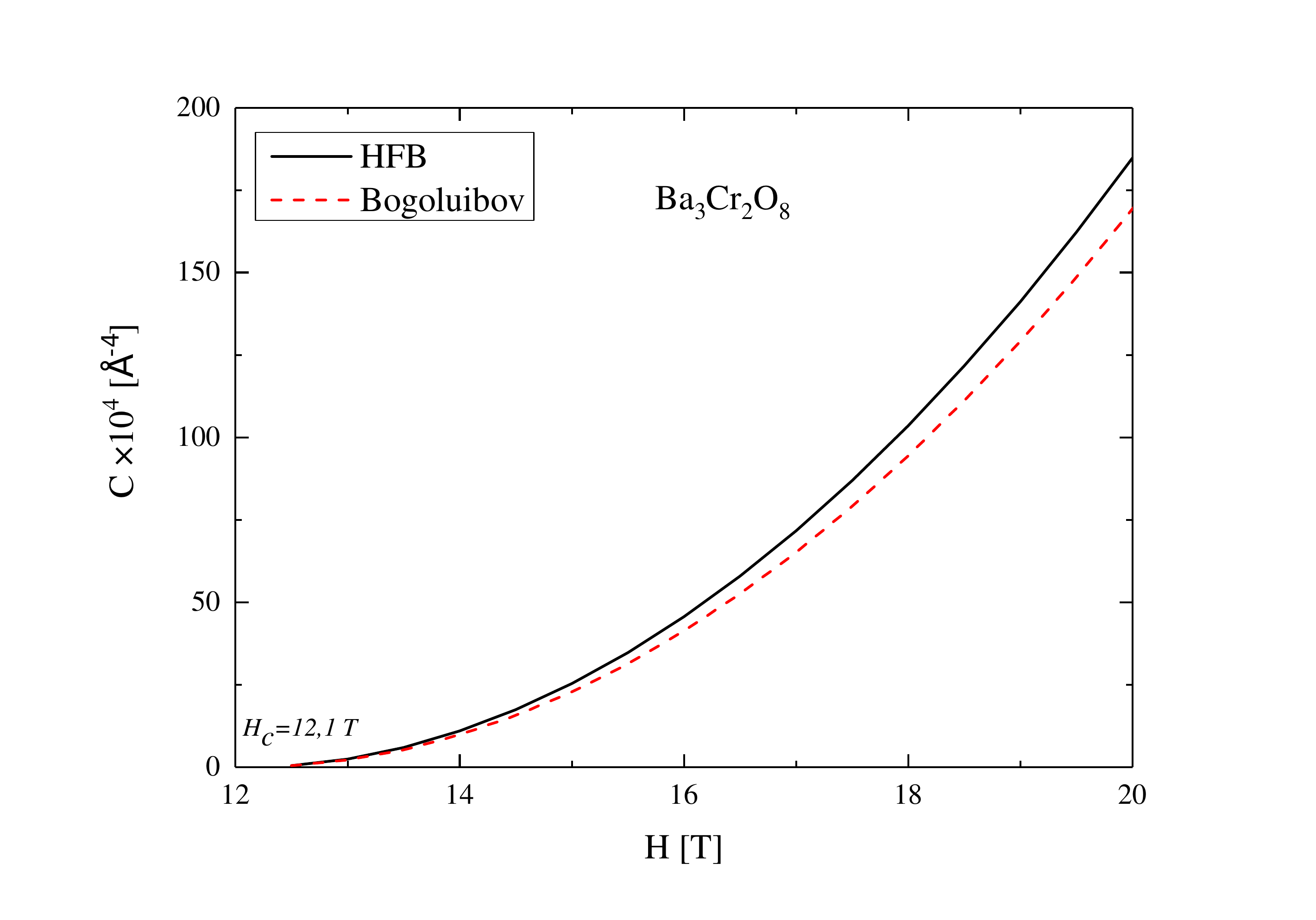} \\ b)}
\end{minipage}
\caption{Tan's contact at zero temperature for TlCuCl$_3$ (a) and Ba$_3$Cr$_2$O$_8$ (b) in \AA$^{-4}$ 
in HFB (solid lines) and Bogoliubov  approximations  (dashed lines). The input parameters are given in Table 1.}
\label{Fig1}
\end{figure}
%%%%%%%%%%%%%%%%%%%%%%%%%END FIG 1

%%%%%%%%%%%%%%%%%%%% TABLE 1 %%%%%%%%%%%%%%%%%%%%%

\begin{table}[h!]%[ tp]%
\caption {Material parameters used in our numerical calculations. The parameters $g$, $H_{c}$, ${\bar a}$ and $\Delta$ are taken directly from the experimental measurements, while $m$ and $U$ are optimized to fit experimental magnetizations (see Ref. [\onlinecite{ourMCE}] for details)}.
\begin{tabular}{ p{3cm} c c c c c c}
 \hline
 \hline
    {}                    &   $g$   & $H_{c} [\rm T]$
&$m [\rm K^{-1}]$&  $U [\rm K]$ &
$\Delta [\rm K]$ & ${\bar a }$ $ [\AA]$ \\
 \hline
 Ba$_{3}$Cr$_{2}$O$_{8}$  &  1.95 &  12.10     & 0.2
&  20     &  15.85 & 3.97 \\
% \hline
 Sr$_{3}$Cr$_{2}$O$_{8}$  &  1.95 &  30.40       &   0.06     & 51.2      &   39.8 & 3.82             \\
% \hline
  TlCuCl$_{3}$  &  2.06 &  5.1       &   0.02   & 315
&   7.1 & 7.93             \\
 \hline
 \hline
   \end{tabular}
   
 \label{tab1}
\end{table}

Although $C_{0}$ in Eq.\,(\ref{Cbog}) is not good at finite temperatures, it may serve as a characteristic scale 
for the contact $C$. Thus, differentiating $\Omega$ in Eq.\,\re{omegBEC} with respect to 
$U$  we obtain the following explicit expression for the contact in HFB approximation
\be
C(T<T_{c})=C_{0}\left\lbrace \widetilde{\rho} - \frac{\widetilde{\rho}_{1}^{2}+\widetilde{\rho^2}-\widetilde{\sigma}^2}{4}+2\rho_c \mu^{\prime}_{\rm eff}\left[ I_3-1-U\widetilde{\sigma}{\sigma}^{\prime}+\frac{\widetilde{\rho}_{0}}{2}+U{\rho}^{\prime}_1 (2-\widetilde{\rho}_1-\widetilde{\rho})\right] \right\rbrace
\label{CSM}
\ee
where $\widetilde{\rho}=\rho/\rho_{c}$, $\widetilde{\rho}_{1}=\rho_{1}/\rho_{c}$, $\widetilde{\rho}_{0}=\rho_{0}/\rho_{c}$, $\widetilde{\sigma}=\sigma/\rho_{c}$, ${\rho}^{\prime}_1=(\partial\rho_1/\partial\mu_{\rm eff})$, ${\sigma}^{\prime}=(\partial\sigma/\partial\mu_{\rm eff})$, 
%$\widetilde{\mu}^{\prime}_{\rm eff}=2\rho_c(\partial\mu_{\rm eff}/\partial U)$,
 ${\mu}^{\prime}_{\rm eff}=(\partial\mu_{\rm eff}/\partial U)=(\widetilde{\sigma}-\widetilde{\rho}_1)
 /[1+2U({\rho}^{\prime}_1-{\sigma}^{\prime})]$ and $\rho_c=\mu /2U$. The explicit expressions for $I_3$, 
 ${\rho}^{\prime}_1$ and ${\sigma}^{\prime}$ are given in the Appendix. Note that, due to the factoring out 
 of $C_0$, the quantity in curly brackets is dimensionless.

Now, using Eqs.\,\re{cdef} and \re{omegBEC}, yields 
\be
\ba
S(T<T_c)=-\ds\sum_k \ln(1-e^{-\beta E_k})+\beta \ds\sum_k E_k f_B(E_k), \\ 
\\
\left( \dsfrac{\partial C}{\partial T}\right) _{T<T_c}=
-2m^2U^2\left(\dsfrac{\partial S}{\partial U}\right)=
8\beta(Um\rho_c)^2\rho_c I_4 {\mu}^{\prime}_{\rm eff}\, , \\
\\
 \left(  \dsfrac{\partial C}{\partial H}\right)  _{T<T_c}= \dsfrac{1}{g\mu_B}\left( \dsfrac{\partial C}{\partial\mu}\right) =-\dsfrac{2m^2U^2}{g\mu_B}\left( \dsfrac{\partial\rho}{\partial U}\right) =\dsfrac{m^2\mu(\widetilde{\rho}-2\rho_c{\mu}^{\prime}_{eff})}{g\mu_B} = -\dsfrac{2m^2U^2}{(g\mu_B)^2}
\left( \dsfrac{\partial M}{\partial U}\right)\, .
\ea
\label{dCSM}
\ee
Unlike in the normal phase, for $T<T_c$ we failed to find a simple relation %%%%%%%BCT1%%%%% 
between susceptibility and $({\partial C}/{\partial\mu})$ similar to Eq.\,(\ref{chi1}).

In the previous subsection we have shown that at zero temperature the contact may be conveniently approximated 
as $C(T=0)\approx C_0= m^2\mu^2$. We now analyze the temperature dependence of $C(T<T_c)$ given in 
Eq.\,(\ref{CSM}) which requires solving the set of Eqs.\,(\ref{MainEqs}).
In Fig.\,2 {a)}, we present $C(T)$ for three compounds Ba$_3$Cr$_2$O$_8$, Sr$_{3}$Cr$_{2}$O$_{8}$, and
TlCuCl$_{3}$ calculated in the HFB approximation (solid curves) using Eqs.\,(\ref{Cbig}) and 
(\ref{CSM}).
It is seen that contact at finite temperature can be at least qualitatively approximated as 
\be
C(T)\approx 2U^2m^2M^2/(g\mu_B)^2\, ,
\label{C0}
\ee (dashed lines) which coincides with the exact one in the normal phase.
In {Fig.\,2 b)}
the magnetizations for the same compounds as a function of temperature are also presented for completeness.

%%%%%%%%%%%%%%%%%%%%%FIGURE 2 %%%%%%%%%%%%%%%%%%%%%%%%%%%%%%%%

\begin{figure}[h!]
\begin{minipage}[H]{0.49\linewidth}
\center{\includegraphics[width=1.25\linewidth]{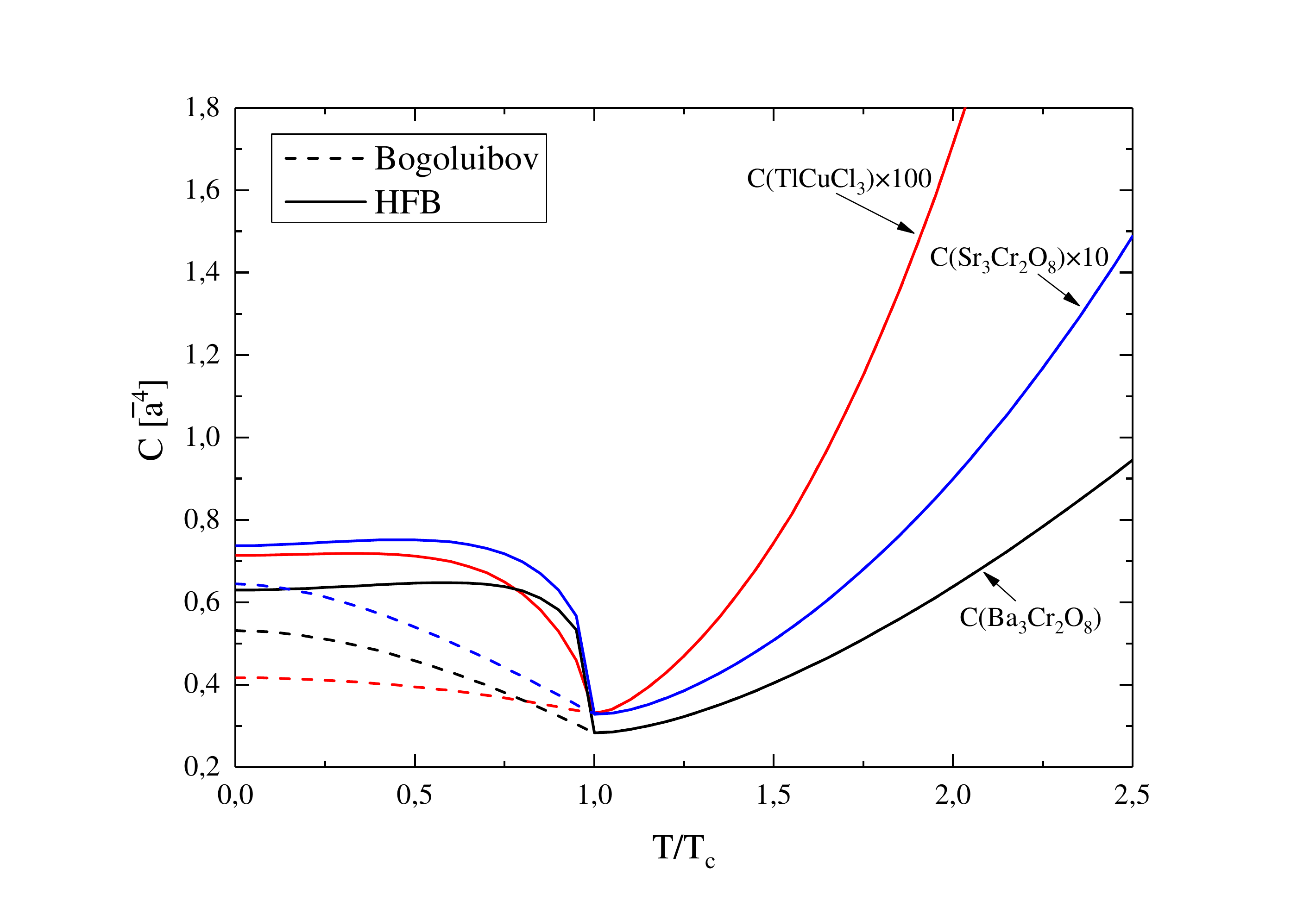} \\ a)}
\end{minipage}
\hfill
\begin{minipage}[H]{0.49\linewidth}
\center{\includegraphics[width=1.25\linewidth]{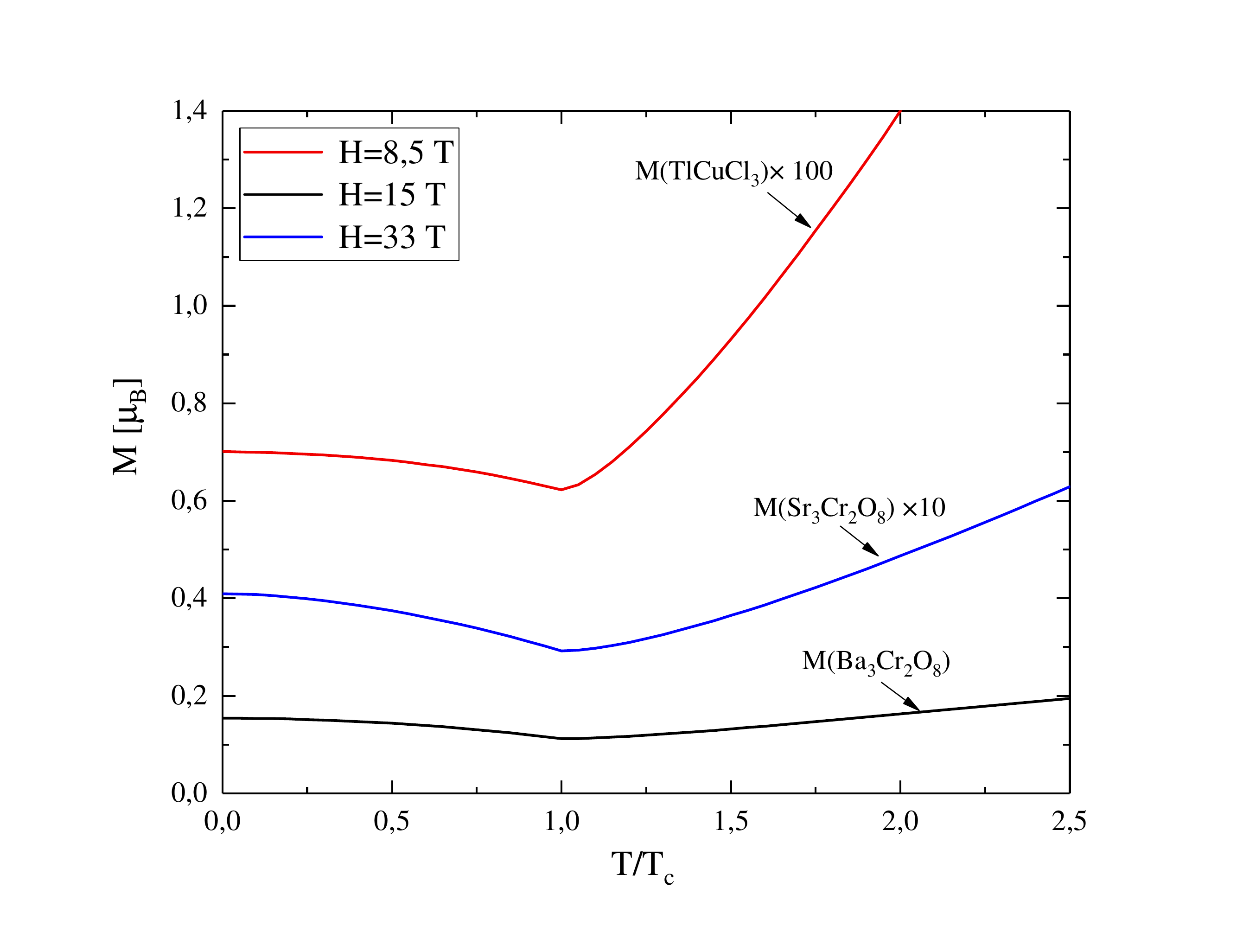} \\ b)}
\end{minipage}
\caption{(a): Contact for spin gapped magnets in dimensionless units. The solid lines are for
HFB approximation. The dashed lines correspond to the classic approximation given in Eq.\,\re{C0}; 
(b): the magnetizations $M(T)$ per dimer.}
%Note that for better description of experimental data on $M(T)$ effects of anisotropies \ci{ourAniz2part2}, which have been neglected here,  should be included.}
\label{Fig2}
\end{figure}
%%%%%%%%%%%%%%%%%%%%%%%%%END FIG 2

%%%%%%%%%%%%%%%%%%%%%   BEGIN FIGURE 3 %%%%%%%%%%%%%%%%%%%%%%%%%%%%%%%%
\begin{figure}[h!]
\begin{minipage}[H]{0.49\linewidth}
\center{\includegraphics[width=1.25\linewidth]{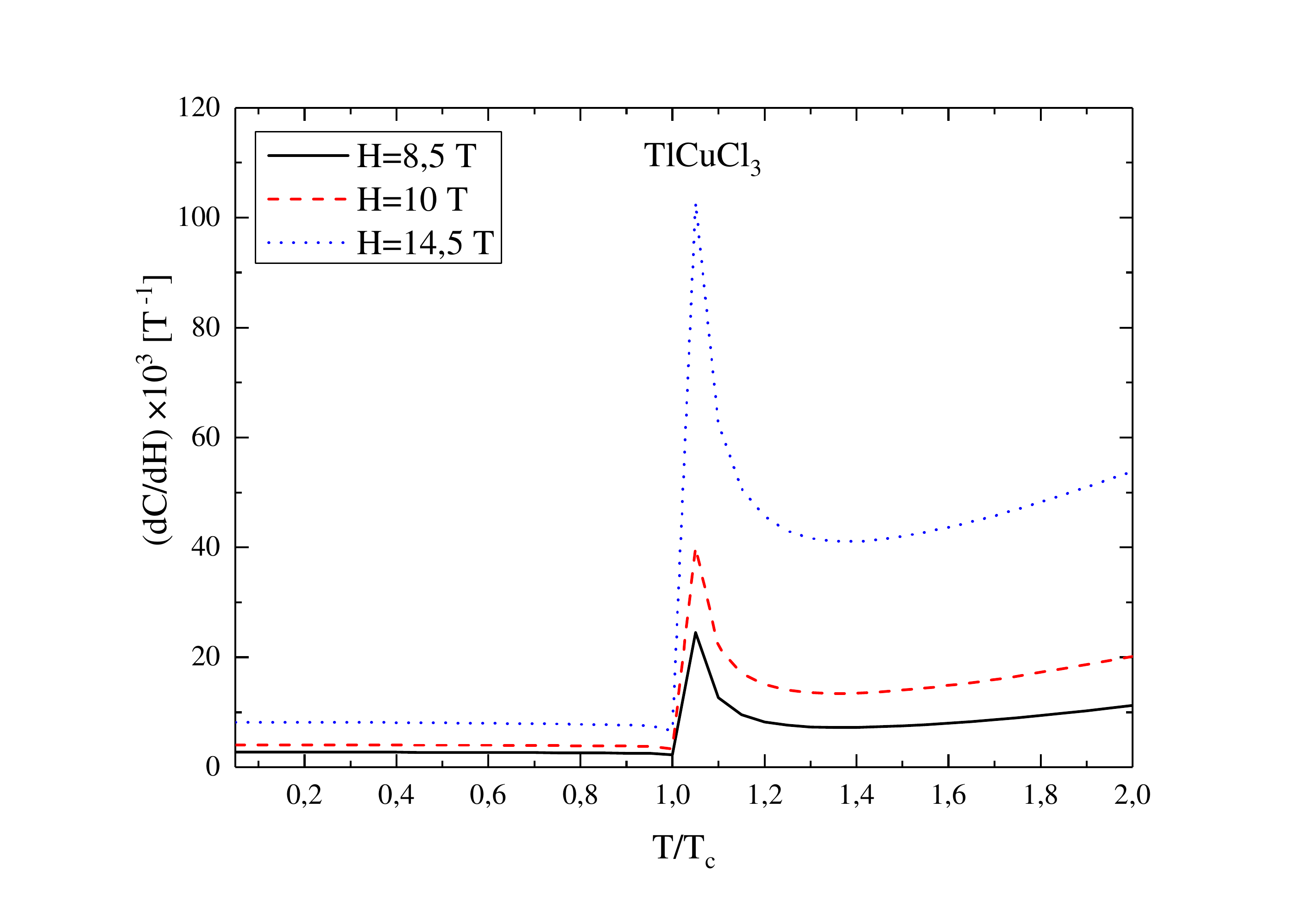} \\ a)}
\end{minipage}
\hfill
\begin{minipage}[H]{0.49\linewidth}
\center{\includegraphics[width=1.25\linewidth]{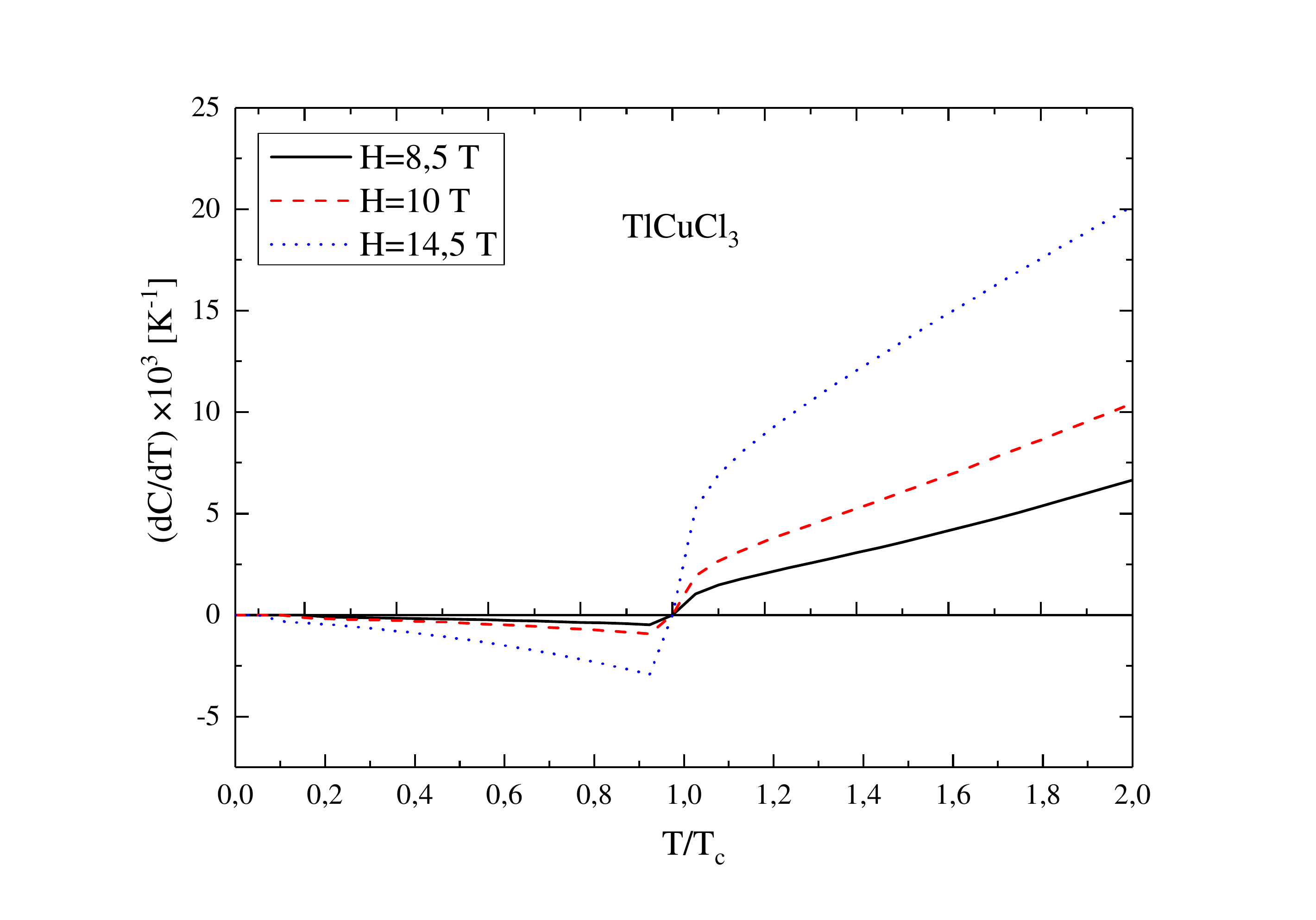} \\ b)}
\end{minipage}
\caption{Derivatives of contact with respect to magnetic field (a) and to temperature
	for TlCuCl$_{3}$. Here we use $V=1$ units, so $(\partial C/ \partial H)$ and $(\partial C/ \partial T)$
	are given in T$^{-1}$ and K$^{-1}$, respectively. The curves are obtained from Eqs.\,\re{dCSM}.
	}
\label{Fig3}
\end{figure}
%%%%%%%%%%%%%%%%%%%%%%%%%%%%  END FIG3

\subsection{Contact and its derivatives near critical temperature}

The crude approximation in Eq.\,(\ref{C0}) is {also useful to explain} the temperature dependence of $(\partial C/\partial H)$ and $(\partial C/\partial T)$ presented in Fig.\,3a and Fig.\,3b, respectively. In fact, it follows that
\bea
\begin{aligned}
\left( \frac{\partial C}{\partial H}\right )\propto
 \left(\frac{\partial M^2}{\partial H}\right )\propto M\chi\, , \\
\left(\frac{\partial C}{\partial T}\right )\propto \left(\frac{\partial M^2}{\partial T}\right )=2 M\left(\frac{\partial M}{\partial T}\right )\, .
\label{propto}
\end{aligned}
\eea
The experimental susceptibilities of quantum magnets with a spin gap show a well-pronounced maximum 
at $T=T_c$,  and they decrease exponentially at low temperatures indicating the existence of a gap $\Delta$ 
between the ground-state and first excited triplet states \ci{delamorre,shermansound,tanaka2020}, while the magnetization 
has a local minimum i.e., $(\partial M/\partial T) |_{T=T_c}=0$ at the critical point (see Fig.\,2b).
This is in good agreement with our predicted temperature dependence of derivatives of the contact: 
$(\partial C/\partial H)$ has a maximum and $(\partial C/\partial T)$ changes its sign near the phase transition from BEC to a normal phase. 

On the other hand, it is well established that the heat capacity $C_V$ has a cusp near the transition point
\ci{ourImanPRE,ourImanIJP,ourAniz2part2,Huangbook}. Therefore, we ask whether the quantities under 
consideration have a jump near $T_c$.
To study this question we consider Eqs.\,(\ref{Cbig}) and (\ref{CSM}). From Eq. (\ref{Cbig}) one may see 
that $C(T\rightarrow T_c^+)=2U^2m^2\rho_{c}^{2}$ . On the other side,  setting in 
(\ref{CSM}) $\tilde{\rho}_1=\tilde{\rho}=1$, $\sigma=0$, $\mu_{\rm eff}=0$ and $E_k=\veps_k$,
one obtains $C(T\rightarrow T_c^-)=m^2\mu^2[1/2+(I_3-1)2\rho_c{\mu}^{\prime}_{\rm eff}]=2U^2m^2\rho_{c}^{2}$,
where we used $I_3(T=T_c)=\sum_k f_B(\veps_k)=\rho_c$  and  $\mu=2U\rho_c$. Therefore, Tan's contact is continuous at the critical temperature: $C(T\rightarrow T_c^-)=C(T\rightarrow T_c^+)=2(Um\rho_c)^2$.
As to its derivatives, using explicit expression for $(\partial C/\partial \mu)$ 
and $(\partial C/\partial T)$ given by Eqs.\,(\ref{difCTCmu}) and (\ref{dCSM}), it can be demonstrated that
they have a cusp at $T=T_c$. Presenting the details in the Appendix, we bring here the final expression for the discontinuities  in dimensionless units 
\be
\ba
%\begin{aligned}
R_{H}\equiv\frac{H_c}{C_{0}}\left\lbrace \left( \frac{\partial C}{\partial H}\right) _{T=T_c-0}-
\left( \frac{\partial C}{\partial H}\right) _{T=T_c+0} \right\rbrace=\frac{4T_c}{U}\left\lbrace \sum_k\left[1+\frac{4f_B(\veps_k)T_c}{\veps_k}-4f_{B}^{2}(\veps_k) \right] \right\rbrace^{-1}\, , \\
\\
R_T=\frac{T_c}{C_{0}}\left\lbrace \left( \frac{\partial C}{\partial T}\right) _{T=T_c-0}-
\left( \frac{\partial C}{\partial T}\right) _{T=T_c+0}\right\rbrace=-\dsfrac{R_H}{\rho_c T_c}\sum_k\veps_k e^{\beta\veps_k}f_B(\veps_k)\, .
\label{RHURT}
%\end{aligned}
\ea
\ee
The numerical values of discontinuities $R_H$ and $R_T$ for TlCuCl$_3$ for some values of the magnetic field are presented in Table 2.
%%%%%%%%%%%%%%%%%TABLE 2%%%%%%%%%%%%%%%%%%%%%%%%%%%%%%%
\begin{table}[h!]%[ tp]%
\caption {Some critical parameters of TlCuCl$_{3}$.}
\begin{tabular}{p{0cm} c c c c c c }
 \hline
 \hline
    & $\ \ \ \ \ H [\rm T]\ \ \ \ $  & $\ \ \ \ \ T_{c} [\rm K] \ \ \ \ \ $ & $\ \ \ \ \ \rho_{c}\ \ \ \ \ $ & $\ \ \ \ \ C (T=T_c)[\rm nm^{-4}] \ \ \ \ \ $ & $\ \ \ \ \ R_{T}\ \ \ \ \ $ & $\ \ \ \ \ R_{\mu}\ \ \ \ \ $ \\ [0.5ex]
 \hline
  & 6.0 & 2.15 & 0.00120 & 0.000384 &  -0.0245 & 0.0134 \\
  & 6.5 & 3.06 & 0.00230 & 0.00140 & -0.0365 & 0.0190 \\
  & 7.0 & 3.73 & 0.00340 & 0.00307 & -0.0458 & 0.0230 \\
  & 7.5 & 4.28 & 0.00450 & 0.00537 & -0.0540 & 0.0263 \\
  & 8.0 & 4.75 & 0.00559 & 0.00831 & -0.0609 & 0.0290\\
  & 8.5 & 5.16 & 0.00669 & 0.0119 & -0.0671 & 0.0314 \\
  & 9.0 & 5.54 & 0.00779 & 0.0161 & -0.0727 & 0.0336 \\
  & 9.5 & 5.88 & 0.00889 & 0.0210 & -0.0779 & 0.0355 \\
  & 10.0 & 6.20 & 0.00999 & 0.0265 & -0.0826 & 0.0373 \\   
  & 10.5 & 6.49 & 0.0111 & 0.0326 & -0.0870 & 0.0390 \\
  & 11.0 & 6.77 & 0.0122 & 0.0394 & -0.0911 & 0.0405 \\
  & 11.5 & 7.04 & 0.0133 & 0.0468 & -0.0950 & 0.0420 \\
  & 12.0 & 7.28 & 0.0144 & 0.0549 & -0.0986 & 0.0420 \\     
 \hline
 \hline
 \end{tabular}
    \label{tab2}
\end{table}
%%%%%%%%%%%%%%%%%%%%%%%%%%%%%%%%%%%%%%%%%%%%%%%%%%%%%%%%%%%%
From Table 2 it may be concluded that the jump in $(\partial C/\partial T)\propto R_T$
is negative ($R_T<0$), in contrast to the jump in $(\partial C/\partial\mu)\propto R_\mu>0$,
and in the heat capacity as well as in the Gr{\"u}neisen parameter \ci{ourMCE}. 
Thus, from the relation $R_T\approx-2m^2U^2(\partial S/\partial U)\vert_{T=T_c}$  one may conclude that,
in $\mu VT$ ensembles entropy increases with increasing repulsive interaction strength, 
i.e., $(\partial S/\partial U)>0$. Next, we discuss another critical phenomenon, which is 
related to the low temperature critical behavior. 

\subsection{Low temperature expansion and QPT}

In a $\mu VT$ ensemble, studied in the present work,  a quantum phase transition occurs at zero temperature 
upon tuning the external magnetic field or equivalently the chemical potential. At $T=0$ the distance to the 
quantum critical point  is determined by a control parameter, $r(H)$. Near the QCP the control parameter 
can be linearized around the critical magnetic field $H_c$ as $r(H)=(H-H_c)/H_c$, which corresponds to 
$\mu=r\Delta$ (where $\Delta=g\mu_B H_c$ is the spin gap).
It is expected that some thermodynamic quantities will diverge at QPT. For example, the Gr{\"u}neisen parameter
which plays an important role in magnetocaloric effect diverges as $\left.\Gamma_{H}\right|_{r\rightarrow 0}\sim 1/r$
\ci{garst,Zhu}. In the following we address the critical behavior of Tan's contact and its derivatives close to QCP. 

The case of $C$ and $\lqavs\partial C/\partial\mu\rqavs$ is rather straightforward. 
Actually, in Section\,II, we have shown that at $T=0$ the contact can be simply approximated as 
$\left. C \right|_{T=0}=m^{2}\mu^{2}$, especially at weak magnetic fields, i.e., as $r\rightarrow 0$ (see Fig.\,1). 
Thus, one directly obtains 
\be
\ba
 C(T=0) \vert_{ \ r\rightarrow 0}=m^{2}r^{2}\Delta^{2}+O(r^{5/2})\, , \\
\left(  \dsfrac{\partial C}{\partial\mu} \right) \vert_{T=0, \ r\rightarrow 0}=2 m^{2}r\Delta+O(r^{3/2})\, ,
\label{QPT1}
\ea
\ee
to show that in contrast to  $\Gamma_{H}$, contact and its derivative  $(\partial C/ \partial H)$ are regular near QCP.

The case of $(\partial C/ \partial T)$ is a little more complicated. Here, one has to find the low temperature expansion 
for $(\partial S/ \partial U)$, as well as small $r$ expansion for the effective chemical potential $\mu_{\rm eff}(r)$. 
This leads to following expressions \ci{ourMCE}
\be
\ba
\left( \dsfrac{\partial S}{\partial U}\right) =-\mu_{\rm eff}^{\prime} \beta^{2}\sum_k \varepsilon_k e^{\beta E_k}f_{B}^2(E_k)=-\dsfrac{\mu_{\rm eff}^{\prime}\pi^2 T^{3}}{15 m s_{0}^5}+O((Tm)^5)\, , \\
\\
\mu^{\prime}_{\rm eff}=\left( \dsfrac{\partial\mu}{\partial U}\right) =-\dsfrac{rQ_{0}m\Delta_{1}}{\pi\Delta}+O(r^{3/2})\, , \\
\mu_{\rm eff}=r\Delta_1+r^{3/2}\Delta_{32}\, ,
\label{Qpir}
\ea
\ee
where $s_0=\sqrt{\mu_{\rm eff}/m}$ is the sound velocity at $T=0$, $\Delta_1=\Delta\pi/(\pi+Q_{0}Um)$, $\Delta_{32}=4(\Delta m)^{3/2}\sqrt{\pi}U/3(Q_{0}Um+\pi)^{5/2}$ and $Q_{0}=(6/\pi)^{1/3}$ is the Debye momentum
\ci{ourMCE}.
Then, the low temperature expansion for $(\partial C/\partial T)$ near QCP can be directly obtained as
\be
\left( \frac{\partial C}{\partial T}\right) =-2m^{2}U^{2}\left(\frac{\partial S}{\partial U} \right)=
-\alpha\frac{T^{3}}{r^{3/2}}+O((Tm)^{5})\, ,
\label{dCdTQCP}
\ee
where $\alpha={m^{9/2}U^{2}\pi Q_{0}}/{15\Delta\sqrt{\Delta_1}}$.
It is seen that at low but finite temperatures $(\partial C/\partial T)$ diverges as $1/r^{3/2}$, but there 
is no divergence at exactly $T=0$ , where QPT occurs. 
Numerical estimations show that $\alpha$ is positive and rather small and hence, at low temperatures the contact decreases very slowly as seen in {Fig.\,2 a).}

%\section{Discussion and Conclusion}

%\newpage
\section{Discussion and Conclusion}

We have studied the Tan's contact for spin-gapped quantum magnets, assuming that their low temperature properties are related to that of a triplon gas.  To the best of our knowledge, this is the first time the parameter $C$ and its dependence on temperature is investigated for a solid.
Naturally,  it will be interesting to compare our results with existing ones in the literature. 
However, there is neither a theoretical nor an experimental study of Tan's contact for $\mu VT$ ensembles at finite temperatures and most of the literature on $C(T)$ concern 1D gases in $NVT$ ensembles.

The temperature dependence of $C$ of harmonically trapped 1D Leib-Liniger Bose gas has been studied by Yao \textit{et al.} \ci{YaoPRE121}.
They have found that contact increases at low temperatures, reaches a maximum at $T=T^{*}$ and then starts to
decrease, i.e., $\left.(\partial C/\partial T)\right|_{T<T^{*}}>0$, $\left.(\partial C/\partial T)\right|_{T=T^{*}}=0$ and 
$\left.(\partial C/\partial T)\right|_{T>T^{*}}<0$. 
From Figs.\,1 and 2, one may note that for a 3D system of bosons the situation is quite the opposite; 
$\left.(\partial C/\partial T)\right|_{T<T_c}<0$ and $\left.(\partial C/\partial T)\right|_{T>T_c}>0$. 
But at the critical temperature $T_c$, in both cases $C(T)$ exhibits an extremum as a function of temperature,
maximum for 1D bosons and minimum for triplons, i.e., in both cases $(\partial C/\partial T)$, changes its sign 
near the critical temperature.
Bearing in mind the relation $(\partial C/\partial T)\sim (\partial S/\partial U)$, one may conclude that at $T=T_c$ the interaction dependence of the entropy displays a maximum.
Note that in the case of 1D bosons this maximum provides a signature of the crossover to the fermionized regime, while in the case of quantum magnets it corresponds just to the point of finite temperature phase transition, where accumulation of entropy occurs \ci{garst}.
  
In the present work the temperature dependence of $C(T)$ at large temperatures $T>T_c$, is found to be 
almost linear. This is to be compared with the results by Vignolo and Minguzzi \ci{Vignolo} 
who obtained the large temperature behavior as $C\sim\sqrt{T}$ for a 1D Bose gas in the Tonks-Girardeau limit.
It would be interesting to study the effect of dimensionality on $C(T)$ in $\mu VT$ ensembles  also
\ci{kosterlic}. 
  
We have  studied the critical behavior of $C$ at the quantum phase transition and found that $C$ and 
$(\partial C/\partial\mu)$ are regular at QCP which is in good agreement with predictions made by 
Chen \textit{et al.} \ci{chencrit}.
 
In conclusion it should be underlined that our analytical expressions for Tan's contact of spin gapped 
compounds are expressed in terms of magnetizations which are directly observable. 
For example, at very low temperatures, $T\ll T_c$, one evaluates $C$ simply from the expression 
$C=m^2\mu^2/{\bar a}^{4}$, where the only parameter, effective mass should be estimated.
Note that, in this region contact of a $\mu VT$ ensemble is almost not sensitive to the strength of
the boson-boson interaction $U$, while the dependence on $U$ of the contact of an $NVT$ ensemble 
is rather strong even in the classical approximation, when $C$ is  given by 
$C(NVT)\approx 16\pi^2a^2\rho^2=U^2 m^2 \rho^2$ \ci{ourTan1,Pitbook14}.  

Finally, we venture to suggest a way to measure $C$ in quantum magnets,
{which can be performed in a similar way as it was done in Ref.\,[\onlinecite{wild2012}] (see Appendix B for
details).}
%We estimate the radio frequency to be of the order of the
%gap, i.e., $\Delta/k_BT\sim 7.5$\,K.}

We hope our work will stimulate more studies, especially experimental ones exploring the universality of
$\mu VT$ ensembles in quantum magnets. {Results presented here will deepen our understanding of the 
connection between short-ranged two-body correlations and magnetic phase transitions in 
antiferromagnetic materials, which 
have the potential to be used in developing the next generation of spintronic devices.}
 
%\newpage
%{\textbf{Acknowledgment}} \\
\acknowledgements
This work is supported by the Ministry of Innovative Development of the Republic of Uzbekistan and 
the Scientific and Technological Research Council of Turkey (TUBITAK) under Grant No.\,119N689. AR
acknowledges partial support from the Academy of Sciences of the Republic of Uzbekistan. BT acknowledges
support from the Turkish Academy of Sciences (TUBA).

%\newpage
\section*{Appendix A}
\def\theequation{A.\arabic{equation}}
\setcounter{equation}{0}

%\appendix

%\section{Explicit expressions for some thermodynamic quantities}
%\label{Appa}
%\begin{center} by Maarten M.S. Solleveld\footnote{Institute for Mathematics, Astrophysics and Particle %Physics, Faculty of Science, Radboud University Nijmegen, P.O. Box 9010 (Heyendaalseweg 135), 6500 GL %Nijmegen, The Netherlands}

%\end{center}

%\numberwithin{equation}{section}

%\setcounter{equation}{0}

In this appendix we provide explicit expressions for the integrals introduced in 
Eqs.\,(\ref{drodSUBg}), (\ref{CSM}) and (\ref{dCSM}):
\bea
\begin{aligned}
 &I_1=-\beta\sum_k\omega_k f_{B}^{2}(\omega_k)e^{\beta\omega_k}, \ \ \ \ \ \
I_2=-U\beta\sum_k f_{B}^{2}(\omega_k)e^{\beta\omega_k}, \\
 &I_3=\rho_{c}^{-1}\sum_k\left(\frac{\varepsilon_k W_k}{E_k}-\frac{1}{2} \right), \ \ \ \ \ \ 
I_4=\beta\sum_k f_B^{2}(\omega_k)\varepsilon_k e^{\beta\omega_k}.
\label{I1234}
\end{aligned}
\eea
where $f_{B}(x)=1/(e^{\beta x}-1)$, $W_k=(1/2+f_B(E_k))$, $\omega_k=\varepsilon_k-\mu+2U\rho$ and $E_{k}^{2}=\varepsilon_k(\varepsilon_k+2\mu_{eff})$.

Equation\,(\ref{CSM}) includes $(\partial\rho_{1}/\partial\mu_{eff})$,  $(\partial\sigma/\partial\mu_{eff})$ and $(\partial\mu_{eff}/\partial U)$. The derivatives with respect to $\mu_{eff}$ can be found by differentiating 
Eqs.\,(\ref{MainEqs}) to give
\bea
\left( \frac{\partial\rho_1}{\partial\mu_{eff}}\right) =\sum_k\frac{\varepsilon_k}{E_{k}^2}\left[  \frac{\mu_{eff}W_k}{E_k}+\frac{(\varepsilon_k+\mu_{eff})W_{k}^{\prime}}{4}\right]\, , \\
\left( \frac{\partial\sigma}{\partial\mu_{eff}}\right) =-\sum_k\frac{\varepsilon_k}{E_{k}^2}\left[ \frac{\mu_{eff}W_{k}^{\prime}}{4}+\frac{(\varepsilon_k+\mu_{eff})W_{k}}{E_k}\right]\, ,
\label{drho1dsig}
\eea
where $W_{k}^{\prime}=\beta(1-4W_{k}^{2})$.

As to $(\partial\mu_{\rm eff}/\partial U)$, it can be obtained by differentiation of both sides of the equation 
$\mu_{\rm eff}(U)=\mu+2U\left[ \sigma(\mu_{eff}(U))-\rho_1(\mu_{eff}(U))\right]$ with respect to $U$ 
and then solving it for $(\partial\mu_{eff}/\partial U)$. This leads to
\bea
\left( \frac{\partial\mu_{\rm eff}}{\partial U}\right) =\frac{\mu_{eff}-\mu}{U D}\, , \\
D=1+2U\left[\left( \frac{\partial\rho_1}{\partial\mu_{eff}}\right) -\left( \frac{\partial\sigma}{\partial\mu_{eff}}
\right)  \right]\, .
\label{dmueffdU}
\eea
Now, we proceed to evaluate $(\partial C/\partial H)=-2m^2U^2g\mu_B(\partial\rho/\partial U)$ and
$(\partial C/\partial T)=-2m^2U^2(\partial S/\partial U)$ near the critical temperature $T_c$. 
%%%%%%% maqola 18, 1 formula left %%%%%%%

%{\bf i)} \textbf{$\mathbf{(\partial S/\partial U)}$ and $\mathbf{(\partial\rho/\partial U)}$ at $\mathbf{T=T_c+0}$.}
From the explicit expression
\be
S(T>T_c)=\sum_k \ln(1-e^{-\beta\omega_k})+\beta\sum_k\omega_k f_B(\omega_k),
\label{STbig}
\ee
where $\omega_k=\varepsilon_k-\mu+2U\rho(U)$, one obtains
\be
\left( \frac{\partial S}{\partial U}\right)_{T=T_c+0}=-\beta^{2}\sum_k e^{\beta\omega_k}\omega_k\omega^{\prime}_{k}f_B^2(\omega_k)\, ,
\label{dSdUTbig}
\ee
where $\omega^{\prime}_k=(\partial\omega_k/\partial U)=2\rho+2U(\partial\rho/\partial U)$ and
$(\partial\rho/\partial U)$ is given by Eq.\,(\ref{drodSUBg}). It is straightforward to show that the integral $I_2$ in 
Eq.\,(\ref{drodSUBg}) is divergent at $T=T_c$, where $\omega_k=\veps_k\sim k^2/2m$, and hence 
$\left.(\partial\rho/\partial U)\right|_{T=T_c}=-\rho/U$. This means that $\omega^{\prime}_k(T=T_c^+)=0$ and 
$\left.(\partial S/\partial U)\right|_{T=T_c^+}=0$. Therefore $\left.(\partial C/\partial T)\right|_{T=T_c^+}=0$ and  
$\left.(\partial\rho/\partial U)\right|_{T=T_c^+}=-\rho/U$.

%{\bf (ii)}\textbf{$\mathbf{(\partial S/\partial U)}$ and $\mathbf{(\partial\rho/\partial U)}$ at $\mathbf{T=T_c-0}$}.
Replacing $\omega_k$ by $E_k$ in Eq.\,(\ref{STbig}) and taking the derivative one finds
\be
\left( \frac{\partial S}{\partial U}\right)_{T<T_c}=-\beta^2\left(\frac{\partial\mu_{\rm eff}}{\partial U}\right)\sum_k\veps_k e^{\beta E_k}f^2_B(E_k).
\ee
Now, taking into account that $\mu_{\rm eff}(T_c)=0$, 
$\rho(T_c)=\rho_c=\mu/2U$, $E_k(T_c)=\veps_k$ and using Eq.\,\re{dmueffdU} we obtain 
\be
\left( \frac{\partial S}{\partial U}\right)_{T=T_c^-}=\frac{\mu\beta^2}{UD}\sum_k\veps_k e^{\beta\veps_k}f^2_B(\veps_k)\,
\ee
which proves the Eq.\,(\ref{RHURT}). The derivative $\left.(\partial\rho/\partial U)\right|_{T=T_c^-}$ may be obtained from 
$ {\rho(U)=(\mu_{\rm eff}+U)/2U} $ as 
\be 
\left( \frac{\partial\rho}{\partial U}\right)_{T=T_c^-}=-\frac{\mu(1+D)}{2U^2D}\, ,
\ee
where $D$ is given by Eq.\,(\ref{dmueffdU}) and $(\partial \mu_{eff})/(\partial U)$ should be evaluated at $T=T_c$.

The expansion of entropy and the effective chemical potential near QCP may be found in the appendix 
of Ref.\,[\onlinecite{ourMCE}].

\section*{Appendix B}
\def\theequation{B.\arabic{equation}}
\setcounter{equation}{0}

{Here we outline a sketch of a possible method of  measuring Tan's contact 
for spin gapped magnets. We presume that this can be achieved in a similar way to that for atomic gases.}

{Wild et al. \cite{wild2012} measured $C$ for $^{85}$Rb atoms in a gas with a 
$60 \%$ condensate fraction. The atoms are in the $|F=2,m_F=-2>$ state, where $F$ is the total atomic 
spin and $m_F$ is the spin projection. Then by applying an rf pulse the atoms are driven from
$|2,-2>$ to $|2,-1>$ state. The rate for transferred atoms to the final spin state is given by \ci{Br9,wild2012}
\begin{equation}
\lim_{\omega\rightarrow\infty}\Gamma(\omega)= \frac{\Omega^2}{4\pi}F(a,\omega,m)C\, ,
\lab{Gamma}
\end{equation} 
where  $\Omega$ is the Rabi frequency, $F(a,\omega,m)$ is a 
given function of scattering length $a$, the frequency $\omega$, and  the  atomic  mass $m$. Tan's contact has been 
extracted from this equation after a direct measurement of $\Gamma$ as a function of $\omega$.}

%%%%%%%%%%%%%%%%%%%%%   FIGURE 4
\begin{figure}[!t]
\begin{center}
\includegraphics[scale=0.4]{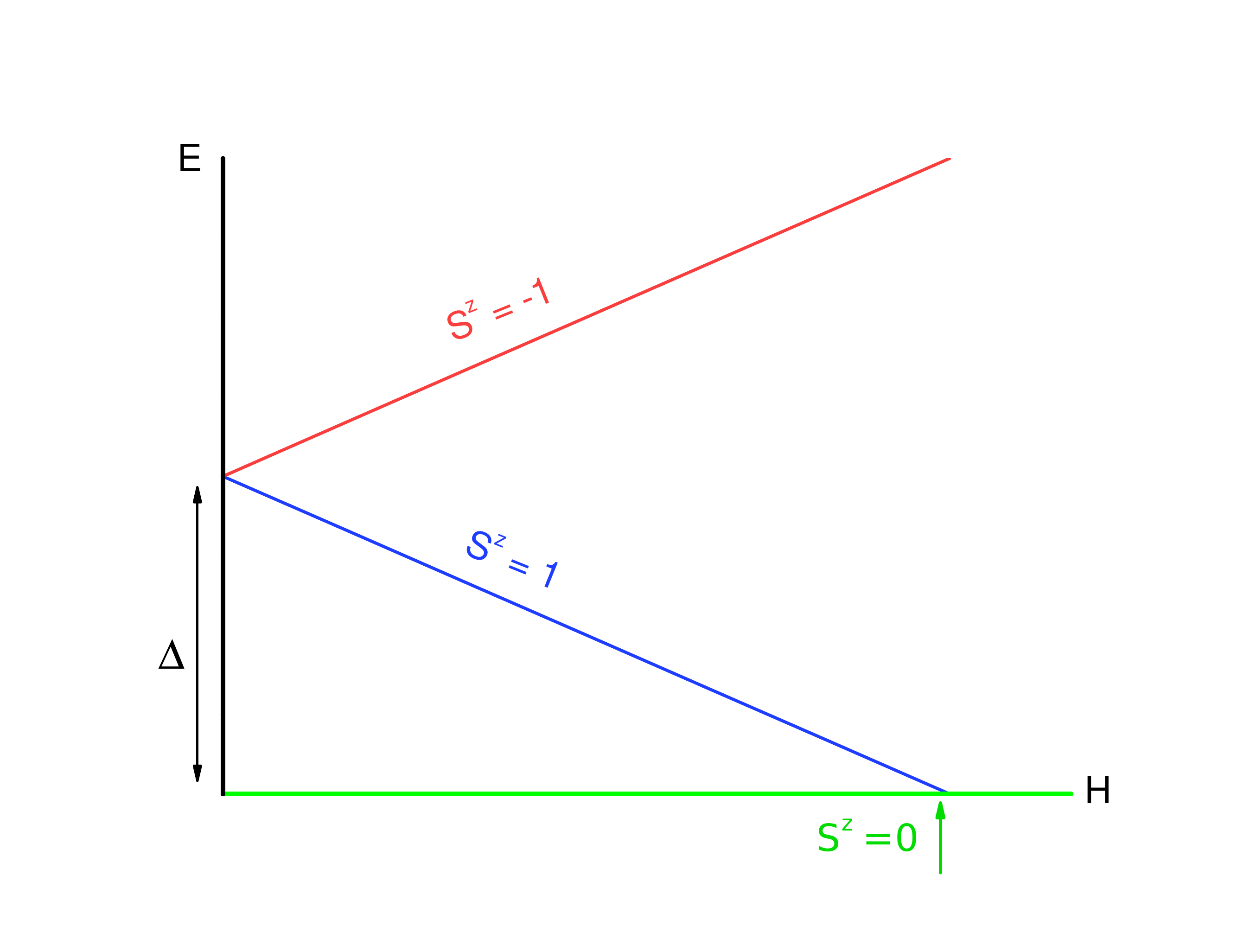}
\end{center}
\caption{Cartoon of the spin levels of a typical spin gapped quantum magnet showing that the 
there are $S_z=\pm 1$ excited doublet and $S_z=0$ ground states. 
They are separated by a gap $\Delta=g\mu_B H_c$, which can be collapsed
due to Zeeman effect for $H>H_c$.
}
\end{figure}

{
Now we consider the spin gapped magnets. Here for $H>H_c$ we have two spin levels with 
$S_z=1$ (ground state) and $S_z=-1$ (nearest excited state), as illustrated in Fig.\,4. The direct transition 
between the spin singlet and triplet states is, in principle, forbidden in magnetic dipole transitions. 
Nonetheless, such transitions have been observed in many spin gapped systems by means of high frequency electron 
spin resonance (ESR) measurements \ci{kimura2004,kimura2018,kimura2020,amr}. 
It was shown that, these transitions are, in fact,  driven
by ac electric fields, and observed electric dipole transitions can be explained by spin dependent polarization.
Very recently, Matsumoto et al. \ci{amr} developed a theoretical description of 
singlet-triplet transitions in spin gapped magnet KCuCl$_3$ in the framework of spin-wave theory,
and proposed practical formulas for the transition probabilty $W$, using Fermi's Golden rule (see e.g.
Eq.\,(18) of Ref.\,[\onlinecite{amr}]). We surmise that if
one uses an effective Hamiltonian similar to Eq.\,\re{Ham} the transition probability from Ref.\,[\onlinecite{amr}]
can be related to the Tan's contact as in Eq.\,\re{Gamma}, and hence one will be able to obtain $C$ from measurements of $W$, using proper normalization of observed quantities.    
}

%\newpage

\end{document}